\documentclass{aa}
\usepackage{epsfig}                  
\usepackage{times}

\newcommand{\BC}{B/C}
\newcommand{\Berat}{$^{10}$Be/$^9$Be}
\newcommand{\Dxx}{D_{xx}}
\newcommand{\gray}{$\gamma$-ray\ }
\newcommand{\grays}{$\gamma$-rays\ }

\def\fwa{80mm}
\def\fha{57mm}
\def\fhb{70mm}

\begin{document}
\hyphenation{an-ti-pro-ton}

   \thesaurus{              
              02.04.2;  
              02.05.1;  
              09.03.2;  
              09.07.1;  
              10.07.1;  
              13.07.3   
              }
   \title{Diffuse Galactic gamma rays, cosmic-ray nucleons and antiprotons}

   \author{I.~V.~Moskalenko\inst{1,2}, A.~W.~Strong\inst{1} \and 
      O.~Reimer\inst{1}}
   \offprints{A.W.~Strong}
   \institute{Max-Planck-Institut f\"ur extraterrestrische Physik,
              Postfach 1603, D-85740 Garching, Germany
   \and       Institute for Nuclear Physics,
              M.V.~Lomonosov Moscow State University,
              119 899 Moscow, Russia
              }
   \date{Received ; accepted }

   \maketitle


   \begin{abstract}
The excess of continuum \gray emission from the Galaxy above 1 GeV is
an unsolved puzzle. It may indicate that the interstellar nucleon or
electron spectra are harder than local direct measurements, as could be
the case if a local source of cosmic rays were to dominate the nearby
flux.  It is however difficult to distinguish between the two cases.
Cosmic-ray secondary antiprotons provide a way to resolve this issue.

We have made a calculation of the cosmic-ray secondary antiproton
spectrum in our model, which computes self-consistently propagation of
primary and secondary nucleons, and electrons.  Fragmentation and
energy losses are computed using realistic distributions for the
interstellar gas and radiation fields, and diffusive reacceleration is
also incorporated.  Our study shows that accurate measurements of the
antiproton flux, especially at high energies, could provide a
diagnostic of the interstellar nucleon spectrum allowing us to test the
hard nucleon spectrum hypothesis. Present antiproton data above 3 GeV
indicate that it can already be excluded at the few $\sigma$ level.

      \keywords{diffusion -- elementary particles -- cosmic rays
                -- ISM: general -- Galaxy: general -- gamma rays: theory
               }
   \end{abstract}

\section{Introduction}
The spectrum of Galactic \grays as measured by EGRET shows enhanced
emission above 1 GeV in comparison with calculations based on locally
measured proton and electron spectra assuming the same spectral shape
over the whole Galaxy (\cite{Hunter97}; \cite{Gralewicz97};
\cite{Mori97}; \cite{StrongMoskalenko97}; \cite{MoskalenkoStrong98a}).
The \gray observations therefore indicate that their spectra on the
large scale in the Galaxy could be different.  Harder cosmic-ray (CR)
spectra could provide better agreement, but the \gray data alone cannot
discriminate between the $\pi^0$-decay and inverse Compton explanations
(\cite{MoskalenkoStrong98b}).  Although the hard electron spectrum
hypothesis seems to be more likely due to the probably clumpy
distribution of electrons at high energies (e.g.,
\cite{PohlEsposito98}), the hard nucleon spectrum cannot be ruled out.
Explicitly, we consider the case that the local nucleon spectrum is not
representative of the regions within a few kpc of the sun, as
could occur if a nearby source of cosmic rays dominates the observed
fluxes.

An important clue may be provided by secondary antiprotons in Galactic
CR produced in collisions of CR particles with interstellar
matter\footnote{Secondary origin of CR antiprotons is basically
accepted, though some other exotic contributors such as, e.g.,
neutralino annihilation (\cite{Bottino98}) are also discussed.}.  These
are an important diagnostic for models of CR propagation and provide
information complementary to that provided by secondary nuclei such as
$Be$, $B$, and heavier nuclei. However, unlike secondary nuclei,
antiprotons reflect primarily the propagation history of the protons,
the main CR component.  The observed intensities depend on the spectrum
of CRs, their composition, details of the nuclear cross sections, and
propagation in the Galaxy. Because they are secondary, antiprotons
reflect the large-scale nucleon spectrum independent of local
irregularities in the primaries.

Previous calculations of secondary $\bar{p}$'s have been made on the
basis of the leaky box model (e.g., \cite{GaisserSchaefer92};
\cite{SimonHeinbach96}) and the locally observed nucleon spectrum.
Recently several experiments have provided improved data on both the
$\bar{p}/p$ ratio and the $\bar{p}$ spectrum itself
(\cite{Hof96}; \cite{Mitchell96}; \cite{Boezio97}; \cite{Moiseev97}),
and the latest calculations by Simon et al.\ (1998) indicate
good agreement with the data.

We have developed a propagation code which aims to reproduce
self-consistently observational data of many kinds related to CR origin
and propagation: direct measurements of nuclei, electrons and
positrons, $\gamma$-rays, and synchrotron radiation. These data provide
many independent constraints on any model and our approach is able to
take advantage of this since it must be consistent with all types of
observation (\cite{Strong96}; \cite{StrongMoskalenko97};
\cite{MoskalenkoStrong98a}).
In this paper we present results on the evaluation of the $\bar{p}$
spectrum and $\bar{p}/p$ ratio in a model including diffusion and
reacceleration and different nucleon injection spectra. Our aim is to
show that $\bar{p}$'s provide a critical test of the alternative
explanations of the GeV \gray excess.  Other secondaries, such as
positrons, also provide a test (\cite{MoskalenkoStrong98b}), but are
more affected by energy losses.
In \cite{MoskalenkoStrong98a} we considered the positron fraction as
evidence favouring a hard nucleon spectrum, but the spectrum considered
was not as hard as required to reproduce the \gray data, and also
absolute positron fluxes were not available at that time. In
\cite{MoskalenkoStrong98b} we show that positron results indeed confirm
the conclusion of the present paper.

\section{Description of the models \label{Description}}
The models are three dimensional with cylindrical symmetry in the
Galaxy, and the basic coordinates are $(R,z,p)$, where $R$ is
Galactocentric radius, $z$ is the distance from the Galactic plane, and
$p$ is the total particle momentum.
The propagation region is bounded by ($R_\mathrm{h}$, $\pm z_\mathrm{h}$)
beyond which free escape is assumed. We take $R_\mathrm{h}=30$ kpc,
$z_\mathrm{h}=4$ kpc since this is consistent with our \BC\ and
\Berat\ study (\cite{StrongMoskalenko98a};
\cite{StrongMoskalenko98b}).  For a given $z_\mathrm{h}$ the diffusion
coefficient as a function of momentum is determined by \BC\ for the
case of no reacceleration; if reacceleration is assumed then the
reacceleration strength (related to the Alfv\'en speed, $v_\mathrm{A}$)
is constrained by the energy-dependence of \BC\ (\cite{SeoPtuskin94}).
The spatial diffusion coefficient for the case of no reacceleration is
taken as $\Dxx = \beta D_0(\rho/\rho_0)^{\delta_1}$ below rigidity
$\rho_0$, $\beta D_0(\rho/\rho_0)^{\delta_2}$ above rigidity $\rho_0$.
The spatial diffusion coefficient with reacceleration is $\Dxx = \beta
D_0(\rho/\rho_0)^\delta$ with $\delta=\frac{1}{3}$ for all rigidities,
and the momentum-space diffusion coefficient $D_{pp}$ is related to
$\Dxx$ (\cite{Berezinskii90}; \cite{SeoPtuskin94}).  The injection
spectrum of nucleons is assumed to be a power law in momentum. The
values used are $D_0=3.5\times10^{28}$ cm$^2$ s$^{-1}$, $\rho_0=5$ GV,
$\delta_1=-0.60$, and $\delta_2=+0.60$ for nonreacceleration models,
and $\Dxx=6\times10^{28}$ cm$^2$ s$^{-1}$ at 3 GV and $v_\mathrm{A}=20$
km s$^{-1}$ for reacceleration models.

The interstellar hydrogen distribution uses HI and CO surveys and
information on the ionized component; the Helium fraction of the gas is
taken as 0.11 by number.  Energy losses for electrons and nucleons are
included (\cite{StrongMoskalenko98b}).
The distribution of CR sources is chosen to reproduce the CR
distribution determined by analysis of EGRET \gray data
(\cite{StrongMattox96}).  The secondary nucleon  source functions are
computed from the propagated primary distribution and the gas
distribution.  The \gray emission from $\pi^0$-decay, inverse Compton
and bremsstrahlung are computed explicitly in 3D from the propagated
nucleon and electron spectra.

\begin{figure}[tb]
   {      
      \psfig{file=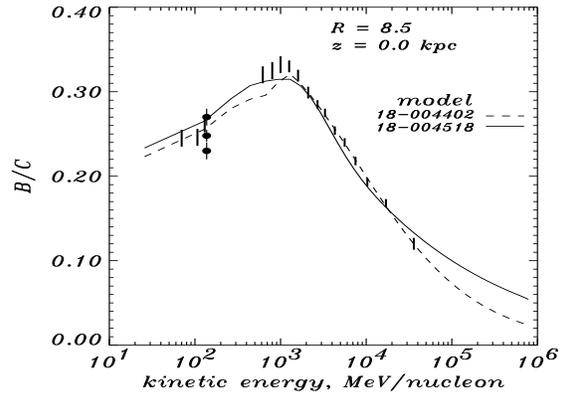,width=\fwa,height=\fha,clip=}
   }
   \caption[fig1]{ \footnotesize
\BC\ ratio for the models with (solid) and without reacceleration
(dashed), $\Phi=500$ MV.  Data:  vertical bars:  HEAO-3, Voyager
(\cite{Webber96}), filled circles:  Ulysses (\cite{DuVernois96}).
\label{f1}}
    \end{figure}

The calculated \BC\ ratio is shown in Fig.\ref{f1} together with recent
data, and the agreement indicates that our propagation models are
satisfactory.
Our preliminary results were presented in \cite{StrongMoskalenko97} and
full results for protons, Helium, positrons, and electrons in
\cite{MoskalenkoStrong98a}.  Evaluation of the \BC\ and \Berat\ ratios,
evaluation of diffusion/convection and reacceleration models, and
limits on the halo size, as well as full details of the methods are
summarized in \cite{StrongMoskalenko98b}.
More details are available on the WWW
(http://www.gamma.mpe--garching.mpg.de/$\sim$aws/aws.html).

\subsection{Antiproton cross sections \label{cross_sections}}
We have used a `standard' formalism to calculate $\bar{p}$ production
and absorption in the interstellar medium. Antiproton production in
$pp$-collisions has been calculated using the Tan \& Ng (1983a)
parametrization of the invariant $\bar{p}$-production cross section.
The total $\bar{p}p$ inelastic cross section has been calculated using
a fit by Tan \& Ng (1983b).  The cross section for $\bar{p}$ production
in proton-nucleus \nolinebreak and nucleus-nucleus interactions has
been obtained (following \cite{GaisserSchaefer92}) by scaling the $pp$
invariant cross section with a factor $F_{it\to\bar{p}X}= (A_i
\sigma_{pt}^\mathrm{inel}+A_t
\sigma_{pi}^\mathrm{inel})/2\sigma_{pp}^\mathrm{inel}$, where $A_{i,t}$
are the atomic numbers of the incident and target nuclei. For the cross
sections $\sigma_{pp}^\mathrm{inel}$ and $\sigma_{pA}^\mathrm{inel}$ we
adapted parametrizations by Tan \& Ng (1983b) and Letaw et al.\ (1983),
correspondingly. The $\bar{p}$ absorption cross section on an arbitrary
nuclear target has been scaled by $A^{2/3}$ using the measured
$\bar{p}\,$--${^{12}C}$ cross section (\cite{Denisov73};
\cite{Carroll79}; \cite{Nakamura84}, \cite{Kuzichev94}).

Simulations of the $\bar{p}$ production with the Monte Carlo
\nolinebreak model DTUNUC (\cite{Simon98}), which appear to be more
accurate than simple scaling, have shown that He nuclei contribute
about 18\% to the total $\bar{p}$ yield and their contribution remains
a constant above the kinetic energy $T_{\bar{p}}\sim 500$ MeV. Heavier
nuclei contribute at about the 3\% level.  Therefore, even if our
simple scaling lowers the $\bar{p}$ yield on nuclei by a factor of 2
(which is  unlikely at $T_{\bar{p}}\ga 500$ MeV), then  the total yield
is not underestimated by more than 10\%. In fact, other uncertainties
dominate the secondary production, for example the form of the
interstellar nucleon spectrum.

Another simplification is that $\bar{p}$'s surviving after an
inelastic collision are totally ignored.  However, calculations made
with only the annihilation cross section show that the difference is
small and the effect can be neglected.

   \begin{figure*}[tbh]
   {      
      \psfig{file=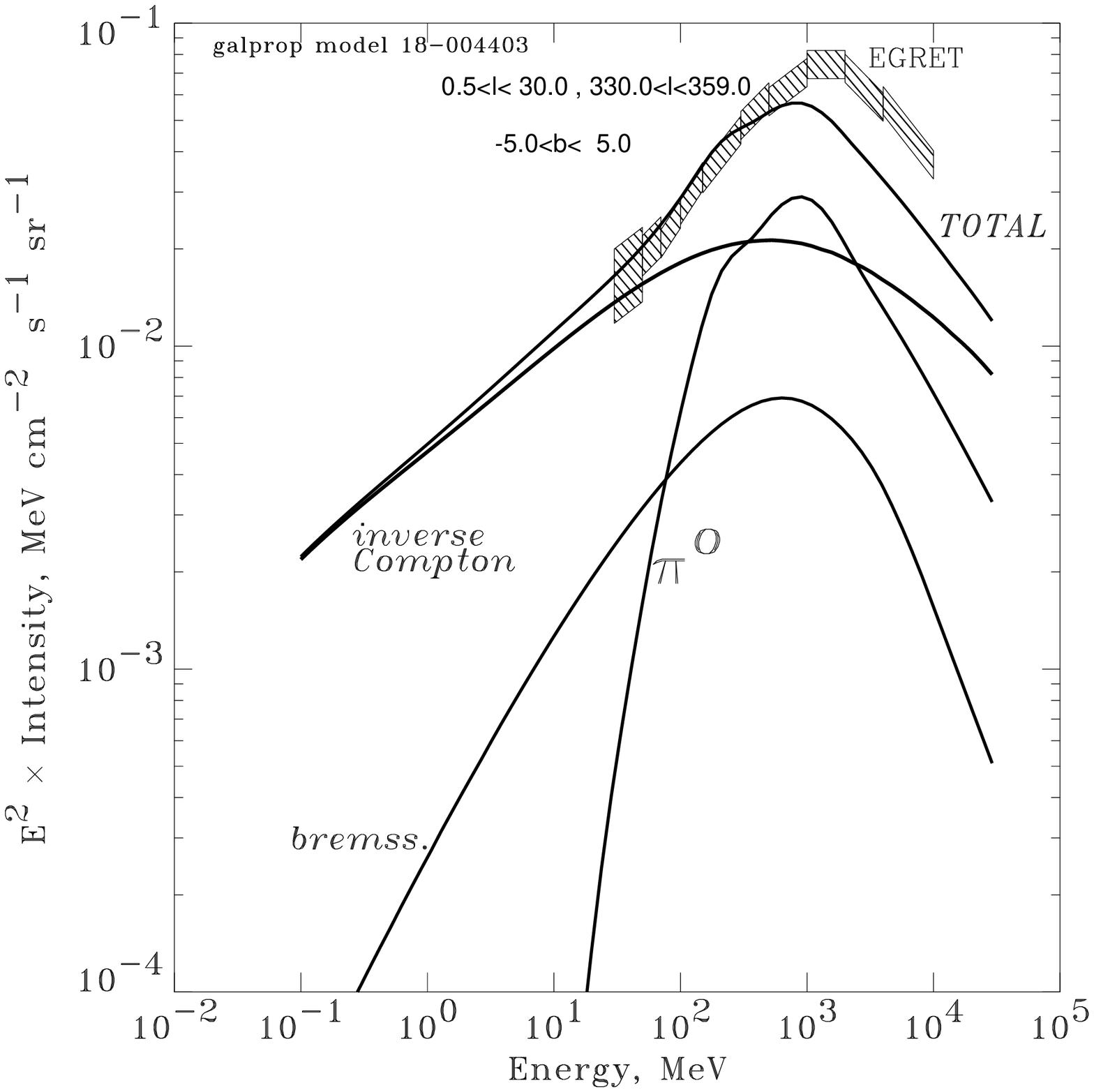,width=\fwa,height=\fha,clip=}
      \psfig{file=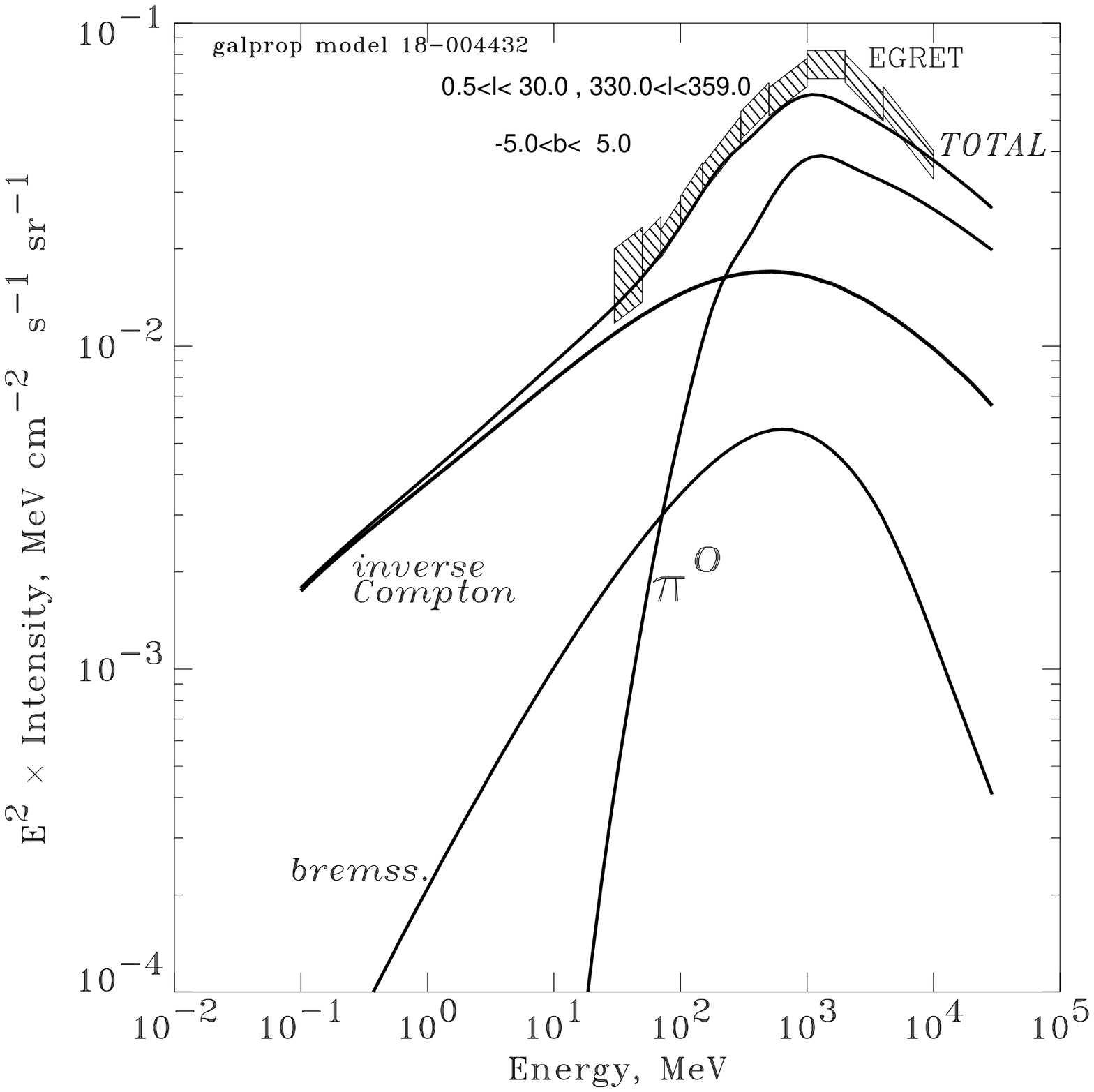,width=\fwa,height=\fha,clip=}
   }
   \caption[fig2a,fig2b]{ \footnotesize
{\it Left panel:} Gamma-ray spectrum of inner Galaxy
($330^\circ$$<$$l$$<$$30^\circ$, $-5^\circ$$<$$b$$<$$5^\circ$) as
measured by EGRET (\cite{StrongMattox96}) compared to model with
`normal' nucleon and electron spectra.  Also shown are the
contributions of individual components:  bremsstrahlung, inverse
Compton, and $\pi^0$-decay.  {\it Right panel:} The same compared to
the model with the {\it hard nucleon} spectrum (no reacceleration).
\label{f2}}
    \end{figure*}

   \begin{figure}[tbh]
   {      
      \psfig{file=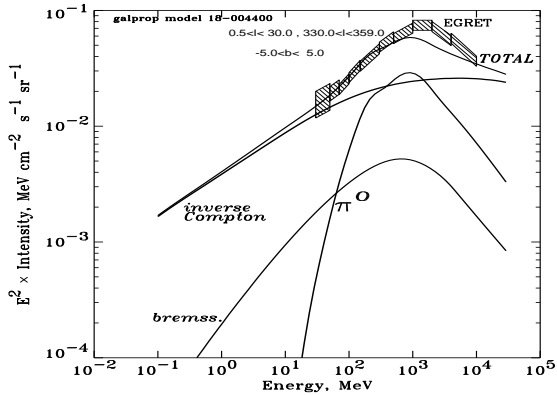,width=\fwa,height=\fha,clip=}
   }
   \caption[fig3]{ \footnotesize
The EGRET \gray spectrum of inner Galaxy compared to model with the
{\it hard electron} spectrum (no reacceleration). Individual components
are the same as in Fig.~\ref{f2}.
\label{f3}}
    \end{figure}

\section{\grays}
Fig.~\ref{f2} (left) shows as an example the \gray spectrum of the
inner Galaxy for a (`normal') model which matches the directly observed
electron and nucleon spectra (the latter is shown in Fig.~\ref{f4} left).
The fit to the EGRET spectra is satisfactory from 30 to 500 MeV and the
deficit above 1 GeV is evident, as discussed in the Introduction.
Simple rescaling of either electron or nucleon spectra does not allow
the agreement to be signficantly improved.

A model with a hard nucleon injection spectrum (no reacceleration,
injection index 1.7) is shown in Fig.~\ref{f2} (right).  The
corresponding propagated interstellar proton spectrum is shown in
Fig.~\ref{f4}.  Fig.~\ref{f3} shows a model with a hard electron
injection spectrum (no reacceleration, injection index 2.0).  Both
models reproduce approximately the observed spectrum, and latitude and
longitude profiles, almost equally well (\cite{MoskalenkoStrong98b}),
and hence it is difficult to discriminate between them.

The same nucleons which contribute to the GeV \gray emission through
the decay of $\pi^0$-mesons also produce secondary $\bar{p}$'s (on the
same interstellar matter).  The harder nucleon spectrum hypothesis,
therefore, can be tested with reliable measurements of CR $\bar{p}$'s.
Above $T_p\sim$ few 10 GeV the mean energy of parent protons is about
10 times larger than the kinetic energy of produced $\bar{p}$'s, and
roughly the same holds for $\gamma$-rays, so 10 GeV $\bar{p}$'s and
$\gamma$'s both are produced by $\approx$100 GeV nucleons. Thus, the
test is well tuned.

\section{Antiprotons}
\begin{figure*}[tbh]
   {      
      \psfig{file=fig4a.ps,width=\fwa,height=\fhb,clip=}
      \psfig{file=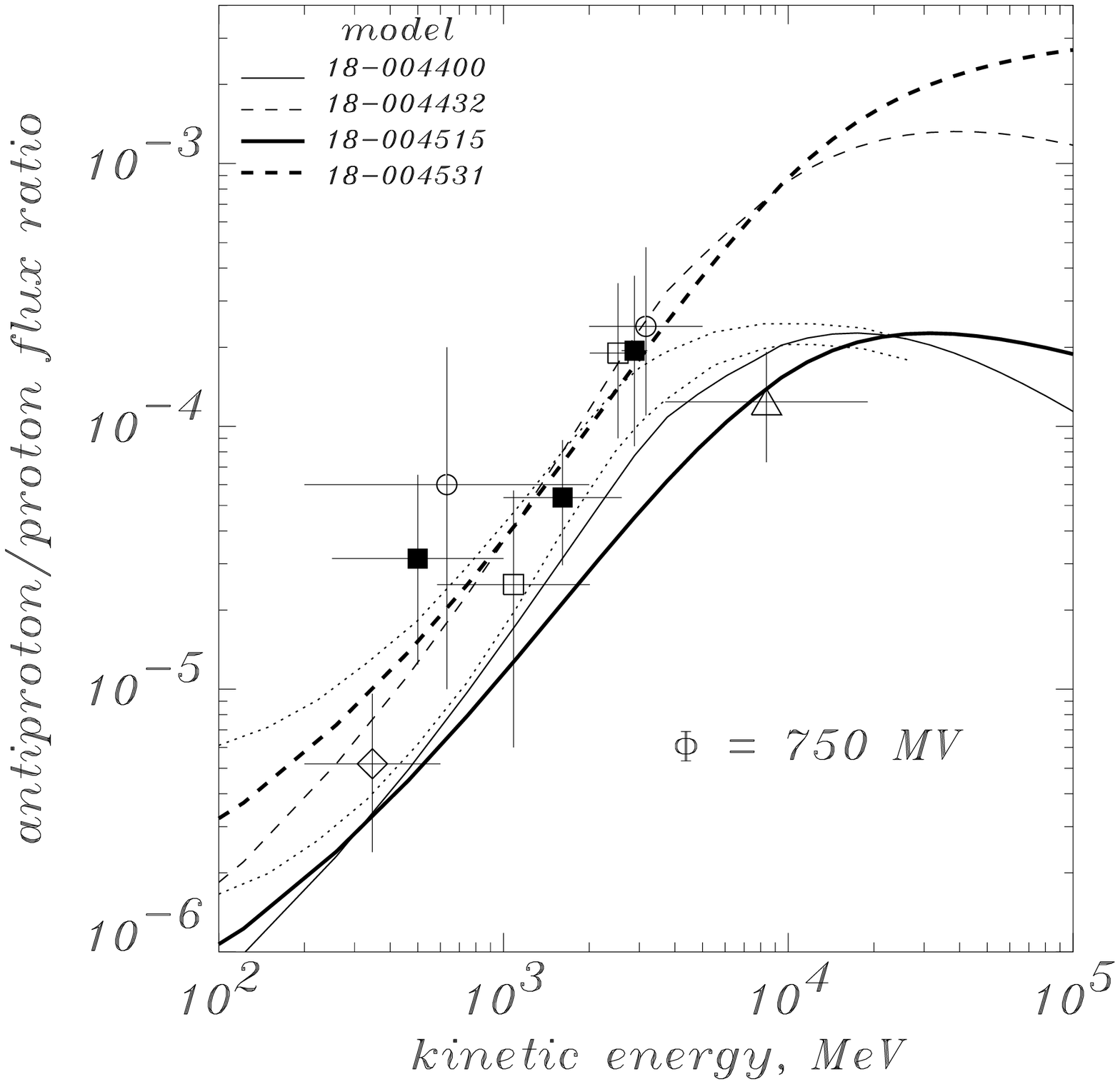,width=\fwa,height=\fhb,clip=}
   }
   \caption[fig4a,fig4b]{ \footnotesize
{\it Left panel:} Interstellar nucleon and antiproton spectra as
calculated in nonreacceleration models (thin lines) and models with
reacceleration (thick lines). Spectra consistent with the local one are
shown by the solid lines, hard spectra are shown by the dashed lines.
The local spectrum as measured by IMAX (\cite{Menn97}) is shown by
dots.  
{\it Right panel:} $\bar{p}/p$ ratio for different ambient proton
spectra. Lines are coded as on the left. The ratio is modulated with
$\Phi=750$ MV.  Calculations of Simon et al.\ (1998) are
shown by the dotted lines. Data from: \rule[0pt]{1.5ex}{1.5ex}\ Boezio
et al. (1997), {\large $\circ$} Bogomolov et al.  (1987,1990),
$\bigtriangleup$ Hof et al. (1996), 
$\sq$ Mitchell et al. (1996), $\diamondsuit$ Moiseev et al. (1997).
\label{f4}}
    \end{figure*}

First we consider the `normal' case, with nucleon injection spectra
which after propagation and modulation match those locally observed
(Fig.~\ref{f4} left).  Our calculations of the interstellar $\bar{p}$
spectra and $\bar{p}/p$ ratio for these spectra are shown in
Fig.~\ref{f4}.  The computed $\bar{p}$ spectrum is divided by the same
interstellar proton spectrum, and the ratio is modulated to 750 MV.
The corresponding ratios are shown on the right panel. We have
performed the same calculations for models with and without
reacceleration and the results differ only in details.  As seen, our
result agrees well with the calculations of Simon et al.\ (1998),
showing that our treatment of the production cross-sections is adequate
as discussed in Section~\ref{cross_sections}.

We now turn to the case which matches the \gray data at the cost of a
much harder proton spectrum than observed (Fig.~\ref{f2} right).  The
dashed lines in Fig.~\ref{f4} (right) show the $\bar{p}/p$ ratio for
the hard proton spectrum (with and without reacceleration); the ratio
is still consistent with the data at low energies but rapidly increases
toward higher energies and becomes $\sim$4 times higher at 10 GeV.  Up
to 3 GeV it does not confict with the data with their very large error
bars.  It is however larger than the  point at 3.7--19 GeV
(\cite{Hof96}) by about $5\sigma$. Clearly we cannot conclude
definitively on the basis of this one point\footnote{We do not
consider here the older $\bar{p}$ measurement of Golden et al.\ (1984)
because the flight of the early instrument in 1979 was repeated in 1991
(\cite{Hof96}) with significantly improved instrument and analysis
techniques. Thus the latter data are more reliable and the relevance of
this measurement to the earlier one is indicated in Hof etal.},
but it does indicate the sensitivity of this test. In view of the
sharply rising ratio in the hard-spectrum scenario it seems unlikely
that the data could be fitted in this case even with some re-scaling
due to propagation uncertainties.  It is interesting to note that the
local $\bar{p}/p$ ratio seems to depend only slightly on the details of
the propagation.

Our main conclusion is that antiprotons provide a {\it sensitive test}
of the interstellar nucleon spectra and hypotheses for the origin of
diffuse Galactic $\gamma$-rays. On the basis of the $\bar{p}/p$ data
point above 3 GeV we seem already to be able to exclude the hypothesis
that the local CR nucleon spectrum differs significantly from the
Galactic average (by implication adding support to the `hard
electron' alternative), but confirmation of this conclusion must await
more accurate data at high energies.
In this respect we note that the $\bar{p}/p$ ratio from Hof et
al.\ (1996) is currently being refined and absolute $\bar{p}$ fluxes
will be calculated (Hof 1998, private communication). Additionally, a
re-flight of the CAPRICE instrument (\cite{Boezio97}) took place in
spring 1998, and several other balloon instruments could be adapted for
antiproton measurements (HEAT:  \cite{Barwick97}), ISOMAX:
\cite{Streitmatter93}). On longer timescale several satellite
experiments are planned or under construction (e.g., PAMELA:
\cite{Adriani95}; AMS:  \cite{Ahlen94}).  These new experiments should
allow us to set stricter limits on the nucleon spectra including less
extreme cases than considered here, and to constrain better the
interpretation of $\gamma$-rays.


\end{document}